\documentclass{sigchi}

\usepackage{balance}  
\usepackage{graphics} 
\usepackage{times}    
\usepackage{url}      
\usepackage{color}
\usepackage{soul}
\usepackage{textcomp}
\usepackage{amsmath,bm}
\usepackage{siunitx}
\usepackage{units}
\usepackage{cite}

\newenvironment{myitem}{\vspace{-0.025in}
\begin{list}{$\bullet$}
{\setlength{\itemsep}{-0pt}
\setlength{\topsep}{0pt}
\setlength{\leftmargin}{10pt}
\setlength{\parsep}{-0pt}
\setlength{\itemsep}{0pt}
\setlength{\partopsep}{0pt}}}%
{\end{list}
\vspace{-0.025in}}

\makeatletter
\def\url@leostyle{%
  \@ifundefined{selectfont}{\def\UrlFont{\sf}}{\def\UrlFont{\normalsize\ttfamily}}}
\makeatother
\def\pprw{8.5in}
\def\pprh{11in}

\usepackage[pdftex]{hyperref}
\hypersetup{
pdftitle={SIGCHI Conference Proceedings Format},
pdfauthor={LaTeX},
pdfkeywords={SIGCHI, proceedings, archival format},
bookmarksnumbered,
pdfstartview={FitH},
colorlinks,
citecolor=black,
filecolor=black,
linkcolor=black,
urlcolor=black,
breaklinks=true,
}

\def\sharedaffiliation{%
\end{tabular}
\begin{tabular}{c}}

\begin{document}

\title{Expanding the Vocabulary of Multitouch Input \\using  Magnetic Fingerprints}

\numberofauthors{3}
    \author{
      \alignauthor Halim \c{C}a\u{g}r{\i} Ate\c{s}\\      
      \email{\url{cagri@cse.unr.edu}}
      \alignauthor Ilias Apostolopoulous \\     
      \email{\url{ilapost@cse.unr.edu}}
      \alignauthor Eelke Folmer\\    
      \email{\url{efolmer@cse.unr.edu}}
      \sharedaffiliation
      \affaddr{Computer Science and Engineering }  \\
      \affaddr{University of Nevada}   \\
       }

\teaser{
\vspace{0.2in}
  \includegraphics[width=7in]{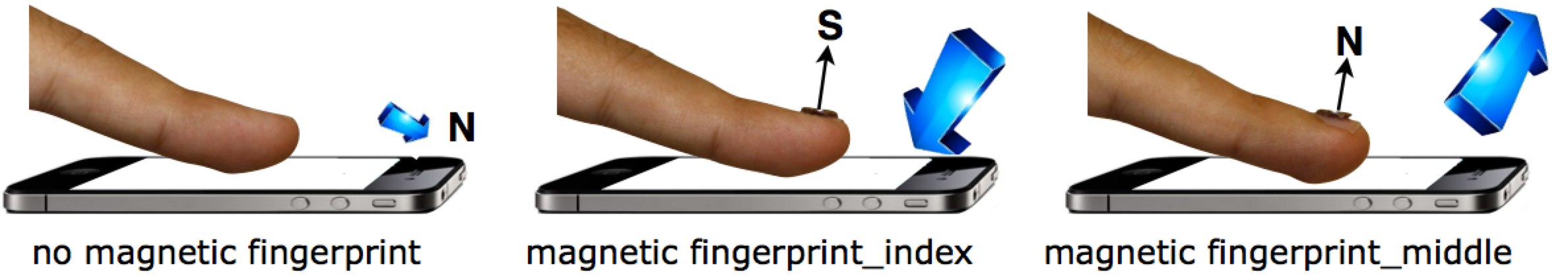}

  \caption{Left: A device's magnetometer (blue arrow) points towards the earth's magnetic field (North). 
Center: A small non-obtrusive magnet is attached to the user's index finger with the South Pole up. When this finger touches the screen, the magnetometer reports the direction and strength of the magnet's field and which allows for identifying the index finger. The magnet's orientation can be used to define different fingerprints as to dinstinguish different fingers. Right: The middle finger is equipped with a magnet with the North Pole up.}
\label{fig:teaser}
}
\maketitle

\begin{abstract}
We present magnetic fingerprints; an input technique for mobile touchscreen devices that uses a small magnet attached to a user's fingernail in order to differentiate between a normal touch and a magnetic touch. The polarity of the magnet can be used to create different magnetic fingerprints where this technique takes advantage of the rich vocabulary offered by the use of multitouch input.  
User studies investigate the accuracy of magnetic fingerprint recognition in relation to magnet size, number of magnetic fingerprints used; and size of the touchscreen. 
Studies found our technique to be limited to using up to two fingerprints non-simultaneously, while achieving a high classification accuracy (95\%) but it nearly triples the number of distinguishable multi touch events. 
Potential useful applications of this technique are presented. 
\end{abstract}


\category{H.5.2.}{HCI}{User Interfaces; Input devices and strategies}

\section{Introduction}

Touch-screens have become a de facto standard of input for mobile devices as they most optimally use the limited input and output space that is imposed by their form factor. 
A single touch event is typically interpreted using spatial information, but precisely manipulating content is often challenging due to small screens, occlusion, variation between users, and finger orientation \cite{Holz:10}; which requires virtual buttons on touchscreens to be larger than the physical buttons they replace. 

To increase the vocabulary of touch interaction, a number of techniques build on or enhance single touch events. Single touch events can be combined into multi-gestures, such as a pinch, but these do not scale up very well due to technical and physiological constraints and do not offer the same accuracy as a single touch input. Temporal features of single touch events can be taken into account to define unique gestures, such as multi-taps or dwelling. 
Different types of touch events can be defined based on the size of a touch event  \cite{Boring:12}. 
Temporal features also can be combined with spatial information, e.g., so called {\it micro-gestures} \cite{Roudaut:09}, but this type of input is considered less efficient and more error prone than when using single touch \cite{Hinckley:05}. 

Beyond counting and timing, a significantly larger touch vocabulary could be created if we could distinguish between different types of single touch events. Some recent work has already explored this idea by creating distinct touch events through the incorporation of additional sensing information acquired with an accelerometer \cite{Hinckley:11} or microphone \cite{Harrison:2011a} (see Related work following) A limitation of these techniques is that they cannot be used as part of multi-touch input, where the largest increase in touch vocabulary may be achieved. 

Though magnets have been previously explored as a mobile input technique those only involve around-the-device interaction \cite{Ketabdar:10,Harrison:09b} or input with a magnet embedded in an object, such as a stylus \cite{Liang:12}. Our approach is novel as we instrument a finger with a small magnet, which allows us to distinguish magnetic fingers from non-magnetic fingers. Our approach requires the smallest amount of user instrumentation. Unlike existing approaches, out technique can be used to augment multitouch input and therefore significantly increase current touch vocabulary. 

\newpage

\section{Related Work}
A number of input technologies are related to our approach. We differentiate approaches that augment single touch input from those that explore using magnets for input. 

{\bf Augmenting single touch input:} The spatial and temporal features of a touch event can be used to define micro-gestures. Microrolls \cite{Roudaut:09}, e.g., unidirectional or circular rolls or rubs made with the thumb can be used to define new types of input, such as opening up a contextual menu. 
Sixteen different micro gestures can be defined with an overall recognition accuracy of 95\%, though this requires per-user calibration.
Thumbrock \cite{Bonnet:13} improves upon the aforementioned approach and achieves a 96\% accuracy in recognizing thumb rolls without calibration.
SimPress \cite{Benko:06} uses the contact area of a touch event generated by the index finger  as a form of simulated pressure to generate different types of single touch events.
Fat thumb \cite{Boring:12} extends this approach to the thumb which has a larger surface area. This approach lends itself well for single hand mobile input where pressing the thumb more firmly to the touchscreen can be used for zooming in and pressing softly for zooming out.  The total number of contact size levels that can be feasibly used or recognized is not investigated. A benefit of these approaches is that they do not require any form of instrumentation though it does require per user calibration. 
Though these approaches work on existing touchscreen devices, a limitation of combining spatial and temporal information of touch events is that recognition is slower and more error-prone \cite{Hinckley:05} than using single touch events.  

Hinckley and Song \cite{Hinckley:11} present a technique that combines capacitive touch sensing with acceleration data, which allows for distinguishing ``hard" from ``soft" taps and swipes. This approach offers a natural semantics as a hard tap can be used to drill down into a menu and a soft tap to go down a single level. No results are presented on the accuracy of this approach, though qualitative results are reported. 
A similar approach is Gripsense \cite{Goel:12} which uses inertial sensors and a vibrator to measure pressure on a touchscreen. When the user touches the screen the vibrator is briefly activated and the damping of vibrations is measured using the accelerometer and gyroscope. This information can then be used to distinguish three levels of finger pressure with high accuracy (95\%) and distinguish different hand postures, e.g., a thumb from an index finger with 84\% accuracy.  Both approaches work on current mobile touchscreen devices without any extra sensors. 

Tapsense \cite{Harrison:2011a} combines capacitive touch sensing with acoustic information acquired with an external high fidelity stethoscope. Capacitive touches made by different objects or parts of the finger make different sounds, which can be classified accordingly. 
For single-handed mobile finger input, four different types of touches made by the tip, pad, nail and knuckle can be recognized with a high 95\% accuracy, but it requires per user calibration.  This technique supports multiuser input, though touch events with different objects cannot overlap. 
Fiberio \cite{Holz:13} a rear projected multitouch table that can optically sense a user's fingerprints. This technique relies on a novel fiber optic plate and unlike magnetic fingerprints, it cannot be used on existing mobile touchscreen devices. Due to the rear projection requirement, this technique is confined to interactive tables.  

A limitation of the aforementioned approaches is that it is challenging to use them as part of a multi-touch gesture made with a single hand. Pressure based approaches only seem to work with a single finger \cite{Hinckley:11,Goel:12}. A thumb roll \cite{Roudaut:09} or a knock with a knuckle \cite{Harrison:2011a} are impractical to use as part of a multi-touch gesture, though this is an area in which the largest increase in touch vocabulary may be achieved. 

{\bf Magnet based input:}  A magnetometer measures the earth's magnetic field, and \--when combined with an accelerometer\-- can be used to determine the absolute 3D orientation of a mobile device. 
Magnetometers are typically used as a compass for navigation or to point out contextual geographic information and are thus widely available in current mobile touchscreen devices. A magnetometer also senses the strength and direction of the magnetic field of a magnet. Because magnets are cheap and do not require an external power source a number of approaches have been explored that use magnets for wireless around-the-device input. 
Abracadabra \cite{Harrison:09b}, Nenya \cite{Ashbrook:11} and MagiTact \cite{Ketabdar:10} use magnet rings to allow for wireless analog input on mobile devices. 
Abracadabra and Magitact use the magnet's distance from the magnetometer to control a cursor or scroll through a list.  By rotating the ring, Nenya uses changes in the direction of the magnetic field to scroll through a list. 
MagiWrite \cite{Ketabdar:10b} enables text input by allowing its users to write digits in the air using a handheld magnet.  

A different set of input techniques embed the magnet into an object. 
GaussSense \cite{Liang:12} embeds a magnet into a stylus, which then facilitates advanced stylus interaction on touchscreens where the orientation and tip pressure of a stylus can be accurately sensed and used for input.  GaussSense requires an external sensor grid that is attached to the back of the mobile device. User studies investigate the accuracy with which finger touches can be distinguished from stylus touches. Magpen \cite{Hwang:13a} is a similar technique that uses a magnetic stylus but works with a device's internal magnetometer. A set of rotary gestures with the pen are proposed. 

Appcessories \cite{bianchi:13} presents a number of tangible objects with magnets embedded in these object. A number of novel interaction options are presented, e.g., objects placed on the touchscreen can be identified using the direction of the magnet's magnetic field. When combined with their spatial location, this information can be used to activate different commands. By rotating an object the change in the direction of the projected magnetic field is interpreted to allow for providing analog forms of input, such as scrolling through a list. 
GaussBits \cite{Liang:2013} is a similar approach that uses objects with magnets embedded in it. GaussBits uses an external magnetic sensor grid and therefore offers a much higher accuracy than when using a device's internal magnetometer. GaussBits supports wireless 3D interaction. MagGetz \cite{Hwang:13b} embeds magnets in physical controllers like switches,  buttons, sliders and joysticks and changes in the magnetic field are interpreted by a device's magnetometer.

\begin{figure*}[t]
\begin{center}
  \includegraphics[width=6in]{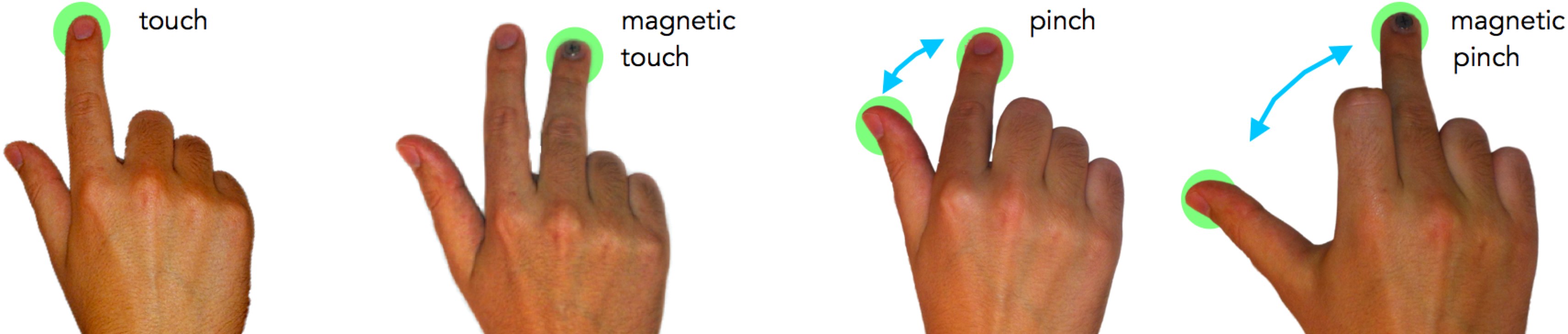}
  \caption{\footnotesize A single magnetic fingerprint doubles the number of distinguishable single and multi-touch events, i.e., we can distinguish two unique single touch events and two different pinch gestures.}
\label{fig:gestures}
\end{center}
\vspace{-0.1in}
\end{figure*}

\section{Design of Magnetic Fingerprints}
We developed a novel input technique, called magnetic fingerprints, that is based on the following observations.
For input, existing magnet input techniques have either explored: 
\begin{myitem}
\vspace{-0.15in}
\item Around-the-device interaction where the user either wears the magnet as a ring \cite{Ketabdar:10,Harrison:09b, Ashbrook:11} or holds a magnet in their hand \cite{Ketabdar:10b}.
\item On-screen interaction either with magnet embedded in an object \cite{Liang:2013,bianchi:13} or stylus \cite{Liang:12,Liang:12b,Hwang:13a,Hwang:13b}. 
\end{myitem}

Rather than using a magnet as a standalone input technique, we use a magnet to {\it augment} finger based touch input. A significant benefit of this approach is that magnetic fingerprints can be used as part of a multitouch gesture, such as a pinch (see Figure \ref{fig:gestures}). Another benefit is that it doesn't require the user to learn new gestures. 
Though current mobile touchscreen devices typically rely on capacitive sensing, magnetic fingerprints could in theory work with any type of touchscreen technology (resistive/optical) as long as it features a magnetometer, which excludes interactive table tabletops. 

A permanent magnet can provide two types of input: (1) the sensed strength of the magnet depends on its distance to the magnetometer and this property can be used to facilitate analog forms of input, such as scrolling \cite{Harrison:09b, Ketabdar:10b}; or (2) the direction of its magnetic field can be interpreted to provide discrete forms of input, i.e., selecting an item \cite{Harrison:09b,bianchi:13}.
A magnetometer reports a 3D vector $ \vec{v}$, which --when there is no magnetic interference-- points towards North $ \vec{n}$ (see Figure \ref{fig:teaser}:left). When a magnet approaches the screen, the vector reported by the magnetometer will change to the value of the magnetic field projected by the magnet $\vec{m}$. 
The direction and magnitude of  $\vec{m}$ differs significantly from the earth's magnetic field, i.e., 
$\vec{m}\neq \vec{n}$ and $|\vec{m}|\gg|\vec{n}|$ (for strong enough magnets).
The strength of the magnet or the orientation of its projected field will generate unique changes in the magnetometer, i.e., $\vec{c} = \vec{n}+\vec{m}$ when the magnet is moved towards the magnetometer. 

It is this property that our approach exploits by instrumenting a fingernail with a small magnet, where the strength of the magnet or the direction of the magnetic field  can be used to define unique ``magnetic fingerprints" that can be distinguished through unique values of $\vec{c}$. 
Various novel input techniques \cite{Yang:12, Chan:13} have explored instrumenting the user's finger with sensors or even a non-obtrusive magnet \cite{Weiss:11}. 
Different approaches could be used for attaching a magnet to a user's finger, such as with a rubber band \cite{Weiss:11}, though it seems most practical for the user to wear a capacitive thimble with a magnet embedded in it. 
Our preliminary experiments showed that wearing a magnetic ring \cite{Harrison:09b,Ketabdar:10} does not allow for accurately distinguishing individual fingers.  
A capacitive glove could accommodate multiple magnets and is available at low cost.

A major consideration in the definition of magnetic fingerprints is its form factor; e.g., the size and the shape of a magnet that needs to be attached to the user's fingernail. 
A magnet's strength depends --among other factors-- on its size. A larger magnet is more easily detected, especially when the magnet is held further away from the magnetometer. For practical and aesthetic reasons it is more desirable to use the smallest possible size magnet.
A challenge with using magnet strength is that different sized magnets may generate the same value for $\vec{c}$ depending on the distance of the magnet from the magnetometer. 
To accurately discriminate different sized magnets, we need to incorporate where the finger touches the screen and a classifier needs to be trained using measured values of $\vec{c}$~for different magnet sizes and locations. 

If we define magnetic fingerprints based on the direction of the magnetic field, the implementation is significantly simpler as we do not need to collect training data and a higher accuracy may be achieved as we only need to look at the direction of change of the components of $\vec{c}$. 
Because this approach doesn't require incorporating where the finger touches the screen, touch less input is one possible application of magnetic fingerprints (see Example Applications). A previous study found that discerning magnets based on their strength is very difficult unless they are really different in size \cite{bianchi:13}, which leads to form factor issues. 
Given the simpler implementation and possible higher accuracy, we only explore defining fingerprints using the direction of the magnetic field.



Because we combine touch sensing with sensing the presence of a magnetic fingerprint, we only require $n-1$ magnetic fingerprints to distinguish $n$ different fingers, as sensing the absence of a magnetic fingerprint in a touch event also defines a unique input event. 
Mobile touchscreens can detect up to five different types of touch gestures, with single touch, two and three finger multitouch gestures being most practical. 
A magnetic fingerprint may be chorded with other non-magnetic fingers to form multi-gestures. 
Using a single magnetic fingerprint, nine unique gestures (two single touch, and seven multitouch gestures) can be defined (see Figure \ref{fig:gestures}). With two magnetic fingerprints, this number theoretically increases to 16 unique gestures, more than three times the amount of what current touchscreen devices are able to detect. Besides practical issues, e.g., fingers could get stuck to each other when magnets attract each other, another potential problem with chording magnetic fingerprints could be magnetic interference. Current magnetometers only report the sum of the complex magnetic field between multiple magnets, which makes it difficult to recognize this gesture. User studies provide insight into how many fingerprints can be used and whether chording is feasible. 


\section{Study 1: Form Factor and Scalability}
For this study, we are interested in understanding how the accuracy of recognizing magnetic fingerprints varies depending on magnet strength and the number of magnetic fingerprints used. A larger magnet is detected more easily but magnetic interference may increase when using multiple magnets. 

\subsection{Instrumentation}
\begin{figure}[t]
\begin{center}
\vspace{-.05in}
\includegraphics[width=28mm]{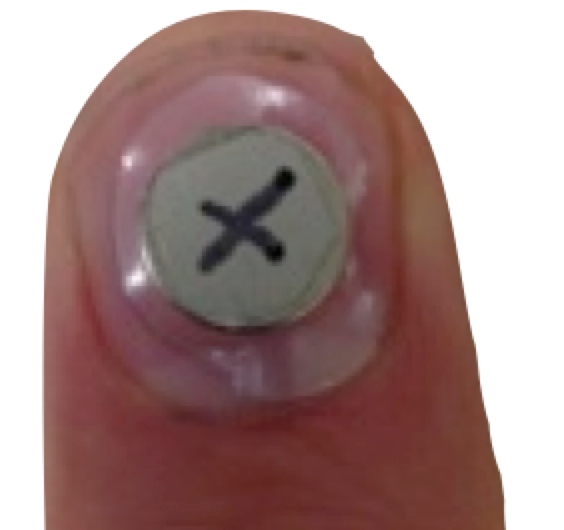}
\hspace{0.2in}
\includegraphics[width=28mm]{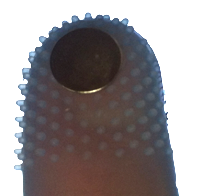}
\caption{Left: for our studies a finger is instrumented with a small neodymium magnet (7.9mm diam.) using an adhesive. 
Right: alternatively the user wears a capacitive thimble with an embedded magnet. 
}
\label{fig:wearing}
\end{center}
\vspace{-0.1in}
\end{figure}
An Apple iPhone 4, with a 1.94" x 2.91" (640 x 960 pixel) display was used for our experiment. 
For this study, we limit ourselves to touchscreen input on smartphones, as an increase in input space offers the largest benefit for this specific platform due to their limited screen real estate. When held in the portrait position, the iPhone's magnetometer is located in the upper right corner. We used N40 grade, disc-shaped Neodymium magnets, which is the strongest type of commercially available permanent magnet at a low cost ($<\$0.10$).
A disc-shaped magnet was chosen as it is least obtrusive and because this shape has been successfully used in a related haptic output technique \cite{Weiss:11}. This shape is more easily attached to a nail than, for example, a cube shaped magnet. Disc magnets are axially magnetized with poles located on the opposing flat circular surfaces of the disc.  Preliminary trials confirmed that the classification accuracy of using magnet strength for defining magnetic fingerprints was much lower than when using the direction of the magnetic field. When using multiple magnets, they need to differ in size significantly as to accurately classify them, which is detrimental to the form factor of this technique. When using the direction of the magnetic field, a magnet of the same size can be used. For these reasons our experiment is limited to evaluating up to two magnetic fingerprints that are defined by the direction of the magnetic field (each pole can only be worn face up). As mobile input is typically limited to single hand input, we deem a potentially three-fold increase in touch vocabulary by exploring the use of two magnetic fingerprints large enough for our study. 
Based on preliminary trials, we decided to explore the following diameters for our magnets (12.7, 7.9 and 3.2 mm with a magnet height of 0.8mm). We marked the north pole of the magnet with an X and attached the magnet to the subject's finger nail using a reusable dot adhesive (see Figure \ref{fig:wearing}:left). 

\subsection{Participants}
We recruited 10 participants (3 female, average age $27.5$, $SD = 3.5$). All subjects were right handed and none had any self-reported non-correctible impairments in perception or impairments in motor control. All subjects were familiar with touchscreen input, as they all owned a smartphone. 

\subsection{Procedure}
The strength of the magnet's magnetic field sensed by the magnetometer varies depending on the distance between the magnet and the magnetometer. To understand what size magnet is required to accurately detect the magnetic fingerprint at every location on the screen, we had subjects perform a single target selection task. 
\begin{figure}[t]
\begin{center}

\includegraphics[width=40mm]{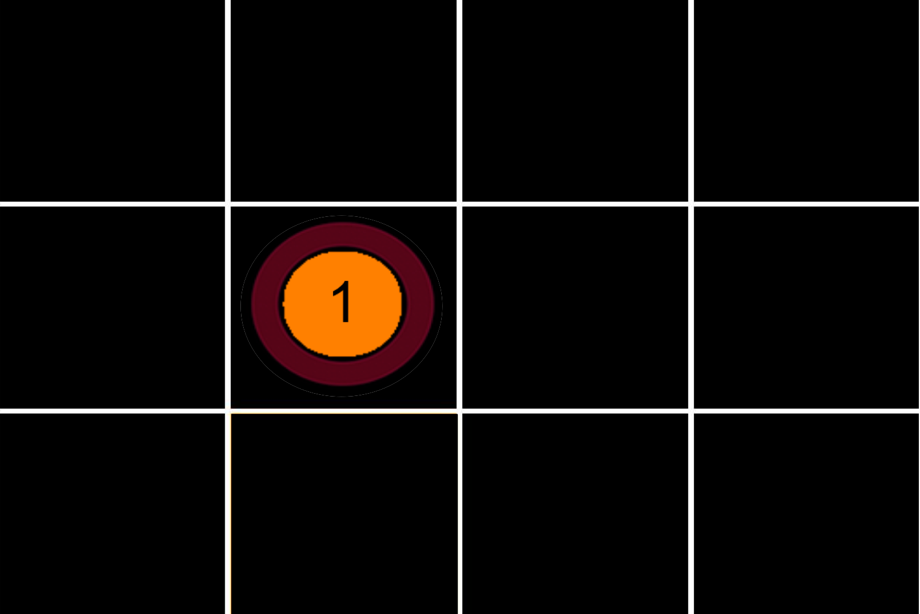}\hspace{2mm}\includegraphics[width=40mm]{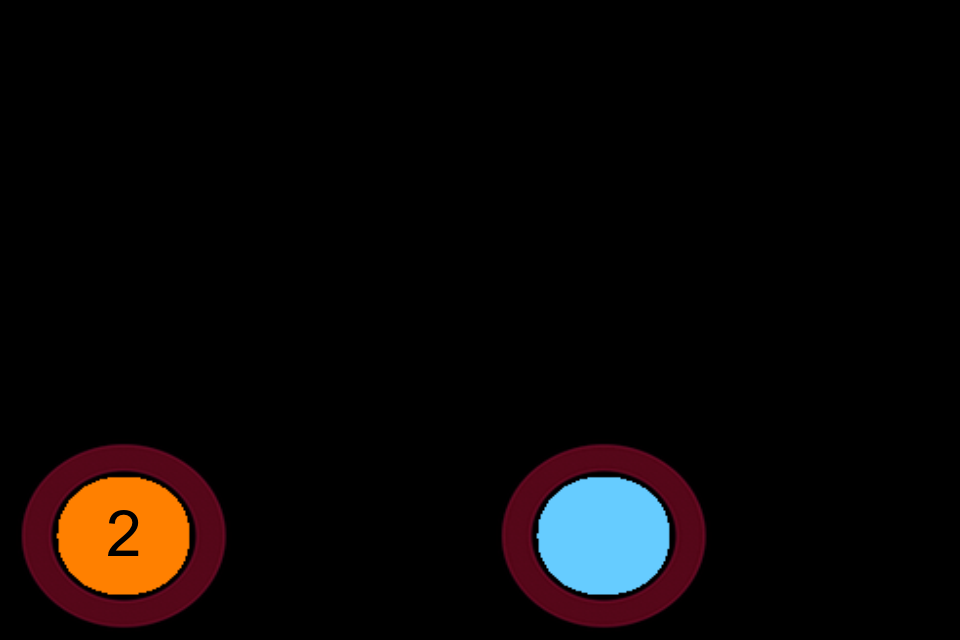}
\caption{Left: the 4x3 grid used for single target selection task.  Right: colored targets that are rendered for the multi-touch selection task.}
\label{fig:grid}
\vspace{-0.1in}
\end{center}
\end{figure}
An iOS application was written that requires subjects to perform a target selection task on a 4x3 grid (see Figure \ref{fig:grid}). 
A target is rendered using an orange circle that measured 220 by 200 pixels (see Figure \ref{fig:grid} left.) 
The (X,Y, Z) values of the magnetometer were recorded in a log file with a sample rate of 40Hz. 
We record the time and location (X,Y) of each touch event.  

We used a repeated measures within-subjects design. Independent variables were magnet {\it size} and {\it number} of magnetic fingerprints.  
For one magnetic fingerprint, we instrumented the middle finger with a magnet (North pole up) and used the middle and index finger for target selection. This setup was used as the index finger is most frequently used for mobile input.
For two magnetic fingerprints, we instrumented the index (North pole up) and ring finger (South pole up) and used these fingers and the middle finger for the target selection task. This setup was chosen to minimize magnetic interference by having the largest possible distance between them. An additional constraint that we identified using a preliminary trial is that better results were achieved for when the user retracts a magnetic finger (place it in the palm) when this finger is not involved in the target selection, as this minimizes false recognitions. Though these specific conditions optimize the use of our technique, it helps establish an upper bound on the best performance. 
We counter-balanced the conditions among subjects. The phone was placed on a flat surface in the landscape orientation and subjects used their dominant hand for target selection. 
Each finger selected 24 targets with two targets per cell in randomized order. Subjects did not switch between fingers, as we were primarily interested in understanding how the accuracy varies depending on where the screen is touched. 

\subsection{Results}
To determine the accuracy with which magnetic fingerprints can be distinguished from each other and from non-magnetic touches, we trained a classifier, i.e., a support vector machine (SVM) using the LibSVM library. To avoid over-fitting our SVM to the training data, we used ten-fold cross-validation using all our data for a total of 480 data points for one magnetic fingerprint and 720 data points for two fingerprints. 
The data was labeled and partitioned into ten equal size subsamples. A single subsample is retained for testing where the other nine subsamples are used as training data. The cross-validation process is repeated 10 times with each of the subsamples used as the validation data, and a single estimate is achieved by averaging the results from each fold. 
Various features were explored to train the SVM, with the best performance achieved for data points containing 62 features, which included: (1) touch location (X,Y); and (2) differential magnetometer data (0.5 seconds) preceding the touch event, which included 20 vectors (X,Y, Z) of  $\vec{c}$ with $\vec{c}= \vec{m} -\vec{v_t}$ where $\vec{m}$ is measured at the touch event $t$ and $\vec{v_t}$ in $25ms$ decrements. Parameters were determined experimentally to yield the best results with no filtering applied. Differential magnetometer values are used as raw magnetometer values vary depending on how the phone is oriented with regard to the earth's magnetic field, which could vary between trials. 

\begin{table}[t]
	\centering
	\setlength{\tabcolsep}{10pt}
	\caption{Classifier accuracy (standard dev)}
	\begin{tabular}{|c||r|r|} \hline
	\bf magnet & \multicolumn{2}{|c|}{{\bf \# magnetic fingerprints}}  \\   \cline{2-3} 
 \bf diameter (mm) & \multicolumn{1}{|c|}{\bf 1}  & \multicolumn{1}{|c|}{\bf 2}  \\\hline \hline 
    12.7 &  98.47 (1.23) & 95.71 (1.68) \\\hline
7.9 & 99.02 (0.95)  & 94.88 (2.73)  \\\hline
 3.2  &  92.18 (3.49) & 80.74 (3.68)  \\\hline
	\end{tabular}
	\label{table:study1}
	\vspace{-0.1in}
\end{table}
Table \ref{table:study1} lists the classification accuracy for each condition --including their standard deviation-- based on the classification accuracy of each fold. 
We performed a two-way ANOVA and found a significant interaction for {\it size} and {\it number} of magnetic fingerprints regarding classifier accuracy $(\bm F_{2,54} =16.926$, $p=.00$, partial $\eta^2=.385$). Post-hoc analysis revealed that 3.2 mm magnet differs in classifier accuracy from the other sizes $(p=.00)$ with no difference between the 7.9mm and 12.7mm magnets ($p=0.99$).  A single magnet yields the best performance $(p=.00$) with an accuracy of 99.02\% (SD=.95) for the 7.9 mm magnet. All errors were due to not correctly recognizing the magnetic fingerprint with 71\% of the errors made by a single subject. 
For two magnets, the best performance is 95.71\% (SD=1.68) using the 12.7mm magnet, with errors uniformly distributed among subjects and types of fingers. Errors were more likely to occur in cells farthest away from the magnetometer. 

\section{Study 2: multi-touch Gesture Recognition}
Where our first study focused on single finger target selection, our second study evaluates the classification accuracy of multi-touch gestures. We focus on two-finger gestures and require subjects to switch between different fingers, which resembles a more practical usage of our technique that we did not evaluate in our first study.
Using a preliminary experiment, we analyzed whether it was feasible to chord two magnetic fingerprints. The thumb and index finger were instrumented with the 12.7mm magnet, which yielded the best performance in our first study. Two fingerprints allow for defining three unique pinch gestures (thumb-index, thumb-middle, index-middle). Three subjects performed a number of different pinches where they had to pinch two targets together. We used fixed targets that were defined around the center of the screen. 
An SVM was trained using the same features as our first study, which yielded an overall classification accuracy of 74.44\% (SD=12.88) with an accuracy of 60.00\% for the pinch gesture involving both magnetic fingerprints. Based on this result, chording magnetic fingerprints does not seem feasible and we limit our study to multi-gestures that use a single magnetic fingerprint at a time. 

\subsection{Instrumentation}
Based on previous results, we used the magnet with a 7.9 mm diameter and we used the same iPhone 4 for our experiment. 

\subsection{Participants}
We recruited eight people (1 female, average age 25.8, SD=4.7) with four people having participated in our first study. All subjects were right handed, owned a touchscreen device and none reported any non-correctable impairment. 

\subsection{Procedure}
We used the same application that we developed for our first experiment, but we modified it so that two targets are rendered that need to be moved together using a pinch gesture (see Figure \ref{fig:grid}:right). For the same reason as our first study, we leave the index finger free and instrument the middle finger with a magnet to allow for defining two unique pinch gestures (see Figure \ref{fig:gestures}). What pinch gesture to provide is indicated using visual cues, e.g., a blue target indicates using a magnetic fingerprint where an orange target a non-magnetized finger. 
Ten unique combinations of targets were defined with one cell distance between the targets. These combinations consisted of six horizontal pinches and four diagonal pinches. 
To establish an upper bound on the performance of this setup, subjects were asked to retract their magnetic finger when it was not involved in the target selection to minimize false recognitions. 
Subjects used their dominant hand for gesture input. For each type of pinch, subjects selected 20 targets for a total of 40 targets with the order of targets and type of gesture being randomized. We also record the number of attempts made for each correct pinch gesture. 

\subsection{Results}
We employed an SVM using ten-fold cross-validation for a total of 160 data points per type of pinch gesture. Using 64 features, e.g., 4 for the touch locations and 60 for differential magnetometer data, we achieved an overall classification accuracy of 97.19\% (SD=3.11) with an accuracy of 96.88\% (SD=4.15) for the non-magnetic pinch and 97.50\% (SD=3.23) for the magnetic pinch, with no significant difference between them  $(\bm t_{18}$ $=.172,$ $p=.693)$.
An analysis of errors found a relatively uniform distribution of errors among subjects, with slightly higher error rate for the cell farthest away from the magnetometer. 
There was no significant difference in classification accuracy with our first study
$(\bm t_{18}$ $=.401$, $p=.093)$, which indicates that switching between fingers for input does not negatively affect performance. 
Users required  an average of 1.12 attempts (SD=.18) for each type of pinch with no significant difference in number of attempts between both pinches  $(\bm t_{7}$ $=.00,$ $p=1.00)$. 
Besides demonstrating the feasibility of using a single magnetic fingerprint to augment multitouch input we found that using the middle finger for a pinch gesture is just as effective as using the index finger.

\section{Study 3: Touchscreen size}
Tablets have become a popular mobile computing platform but they typically feature larger touchscreens than smartphones. For this study, we evaluate whether magnetic fingerprints can be used on tablets. Due to their larger screen size, stronger and larger magnets need to be used. We limit our experiment to evaluating a single magnetic fingerprint. 

\subsection{Instrumentation}
For this experiment, we used the popular Apple iPad 4 tablet, which features 
a 9.50" x 7.31" (2,048 x 1,536 pixel) display (9.7" diagonal). We were unable to verify whether the iPad features the same magnetometer as the iPhone. We used the largest diameter magnet from our first experiment (12.7 mm).  As the 12.7 mm magnet is already at the size of a nail, we explore  magnets with larger height. %
Based on experiments, we evaluated the following heights (0.8, 1.6 and 2.4 mm). 

\subsection{Participants}
We recruited eight subjects (1 female, average age 27.0, SD=4.6). Four subjects had participated in prior studies. All subjects were right handed, owned a smartphone and none reported any non-correctable impairment. 
\begin{table}[t]
	\centering
	\setlength{\tabcolsep}{5pt}
	\caption{Classifier accuracy (standard dev)}
	\begin{tabular}{|c||r|r|r|} \hline
\bf magnet	 & \multicolumn{3} {|c|}{{\bf grid size}}  \\   \cline{2-4} 
 \bf height (mm) & \multicolumn{1}{|c|}{\bf 8 x 6}  & \multicolumn{1}{|c|}{\bf 6\nicefrac{1}{2} x 5} & \multicolumn{1}{|c|}{\bf 4 x 3 }  \\\hline \hline 
  0.8   &  79.40 (4.99) & 81.98 (5.50)  & 93.68 (5.45) \\ \hline
  1.6	 &  81.64 (4.76)  & 81.33 (2.77)  & 93.75 (5.25) \\ \hline
  2.4   &  86.85 (3.70) & 88.11 (4.08) & 98.43 (2.59)  \\ \hline
	\end{tabular}
	\label{table:ipad}
	\vspace{-0.1in}
\end{table}
\subsection{Procedure}
Subjects performed the same target selection task as in our first study. Instead of using a 4x3 grid, we use an 8x6 grid to allow for a comparison of results using a similar cell size (accommodated for differences in screen resolution). Though the iPhone 4 and iPad 4 have slightly different aspect ratios, we did not deem this to be a significant problem. For a single magnetic fingerprint, we instrumented the middle finger with a magnet and used the middle and index finger for target selection. 
Each finger selects 48 targets in the grid in randomized order for a total of 96 targets per magnet size. We counter-balanced the conditions among subjects.

\subsection{Results}
An SVM was trained using the same features as for our first study for a total of 384 data points per type of finger. To understand the effect that touchscreen size has on classification accuracy, we trained SVMs using subsets of our grid. We used a grid of 6\nicefrac{1}{2} x 5 (iPad mini) and 4x3 (iPhone 4) to allow for a comparison with results from the first study. 
Table \ref{table:ipad} lists the results. For the 8x6 grid, a one-way ANOVA found a statistically significant difference in accuracy between magnets $(\bm F_{4,7} =7.108$, $p<.05$) where a post-hoc analysis showed that the 2.4 mm magnet outperforms the 1.6 mm ($p=.049$) and the .8 mm ($p=.00$) magnets. 
An accuracy of 86.85\% (SD=3.70) is achieved for the 9.7" iPad with a slightly higher accuracy (88.11\%) for 8" tablets (iPad Mini). 
Errors were uniformly distributed among subjects and among types of fingers. Similar to our prior results, errors were more likely to occur in cells farthest away from the magnetometer; however, we also observed false recognitions in the cells closest to the magnetometer. 
Though we ensured subjects retracted their magnetized finger, it seems due to the larger screen and the size of the magnet used, false recognitions become difficult to avoid. This also explains the lower accuracy found for the .8 mm magnet for the 4x3 grid. 

\section{Example Applications}
We illustrate the usefulness of magnetic fingerprints using a number of example applications. 

\subsection{Mode Switching}
\begin{figure}[h]
\begin{center}
\includegraphics[width=85mm]{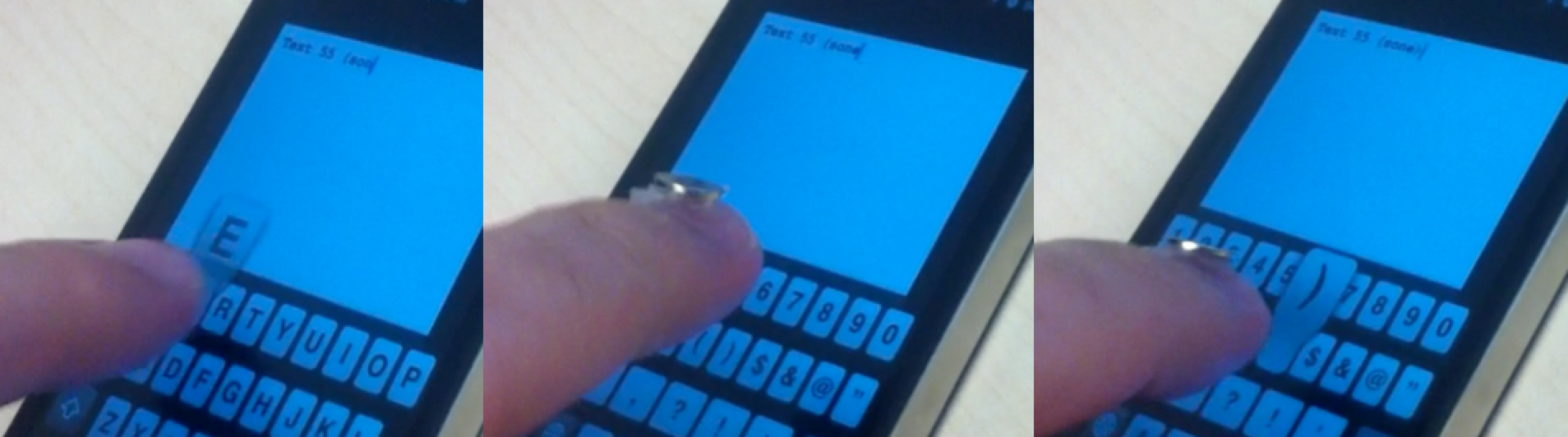}
\caption{Left: regular characters are provided with the index finger and special characters with a magnetic fingerprint. Middle: Before touching the screen, we can sense the presence of a magnetic fingerprint and switch to the numerical keyboard. Right: a special character is selected.}
\end{center}
\vspace{-0.0in}
\end{figure}
Providing special characters on mobile virtual keyboards often requires switching to a special keyboard, which is inefficient as it requires multiple user inputs to select the right key. To address this problem, Harrison et al. presents a solution using their Tapsense technique \cite{Harrison:09b} in which special characters can be provided using different parts of the finger (pad,nail). This approach integrates multiple keyboards into a single keyboard and renders multiple characters on each key, which may cause keys to become illegible. 
Magnetic fingerprints can be used for efficient mode switching. For a mobile keyboard, users type regular characters with their index finger and use a magnetic fingerprint on their middle finger to provide a special character. 
What is novel about this approach is that we can already sense when a magnetized finger is approaching before it touches the screen and switch to the corresponding keyboard. This allows users to hover with their magnetized finger over the keys to find the special character before selecting it. Our approach could support three different keyboards while allowing for legible characters on keys. 

\subsection{Rotary Gestures}
\begin{figure}[h]
\begin{center}
\includegraphics[width=85mm]{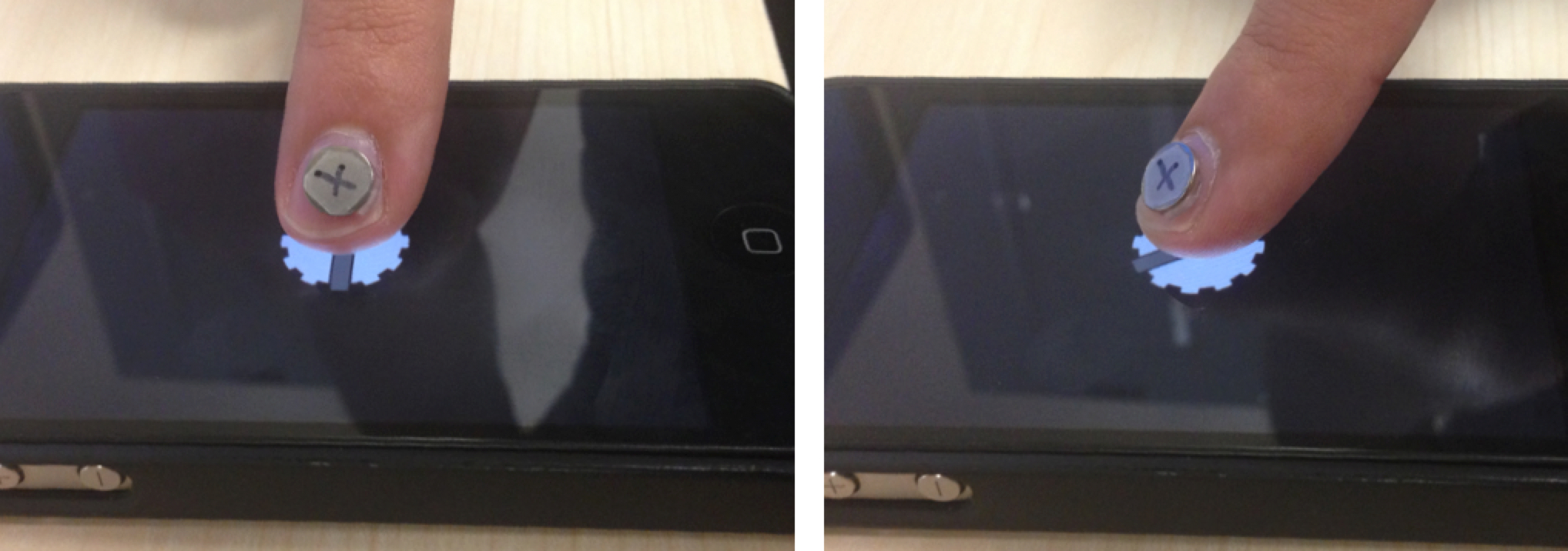}
\caption{Left: a knob is touched with a magnetic fingerprint. Right: When the finger rotates, changes in the magnetic field adjust the knob.}
\label{fig:rotation}
\end{center}
\vspace{-0.0in}
\end{figure}

Knobs or sliders are GUI elements that can be used for adjusting continuous input values, such as volume or pitch in a music application. It is more desirable to use a knob for mobile touchscreen input, as they require less screen real estate than a slider \cite{Gelineck:09}. Knobs are difficult to manipulate using touch as this cannot be achieved with a single touch event and requires interpreting a two finger rotate. Two finger rotates are difficult to perform with a single hand and requires increasing the size of the knob to allow for bimanual access. 
Bianchi et al. \cite{bianchi:13} demonstrates providing rotary input with a magnet embedded in an object. The object is placed on the touchscreen and when it is rotated changes in the direction of its magnetic field are interpreted to adjust a value. A magnet attached to a user's finger can be interpreted in the same way and allows users for to precisely manipulate a knob by rotating their finger instead of an object or using two fingers. Experiments show a feasible input range of nearly $180^{\circ}$. This gesture could also exploit the metaphor or screw/unscrew gestures where a value, e.g., a slider bar handle, can be fixed temporarily.

\vspace{-0.05in}
\vspace{-0.05in}
\subsection{Multi{$^2$}-Touch Gestures}
\begin{figure}[h]
\begin{center}
\includegraphics[width=85mm]{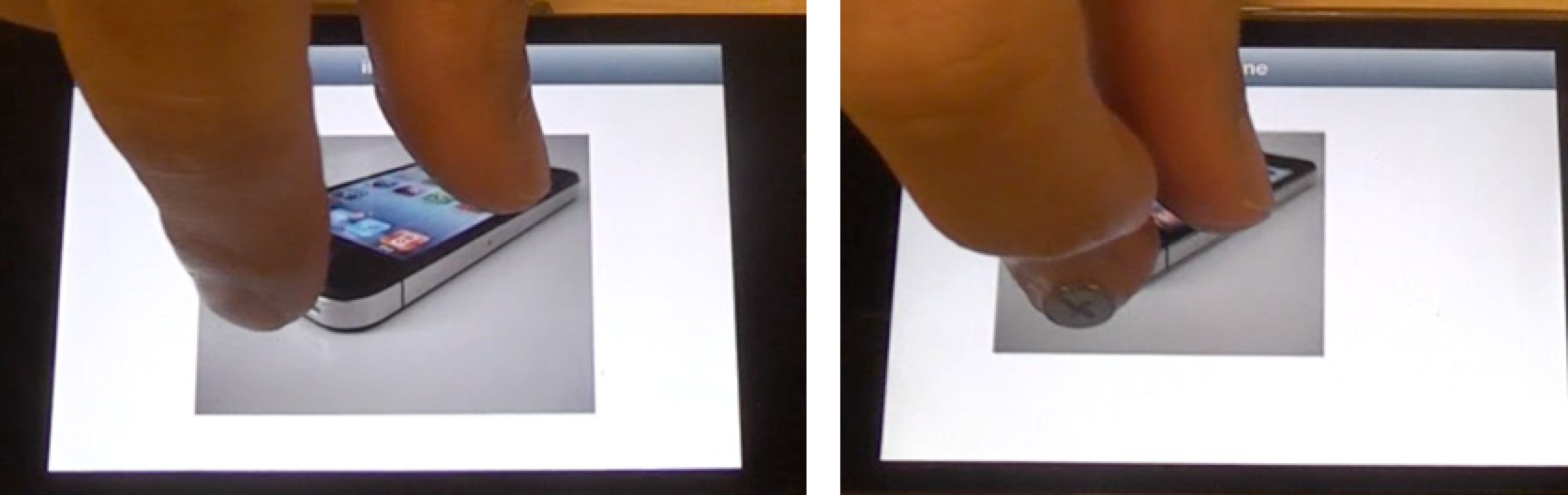}
\caption{Left: a non-magnetized pinch scales a photo. Right: a pinch-in with a magnetic fingerprint picks up the photo where it can be dragged to a new location and a pinch-out drops the photo.}
\label{fig:multipinch}
\end{center}
\vspace{-0.15in}
\end{figure}
Multi-touch gestures, such as pinches, are typically associated with a single type of functionality, such as zooming. In real life, a pinch gesture could be used for different actions, e.g., picking up an object or squeezing an object.  Our study did not find a significant difference in proficiency between making pinches with the index or middle finger. We created a simple mobile photo manipulation application that uses Multi{$^2$}-touch gestures, e.g., multi-touch gestures that afford different types of natural interactions by using different fingers in the gesture. 
A pinch gesture made with the index and thumb is used for scaling a photo but a pinch-in gesture made with a magnetic fingerprint on the middle finger picks up the photo, upon which the user can drag the photo to a new location and drop the photo with a pinch-out.  Though this gesture may be less efficient than using a single finger it doesn't require mode switching and allows for two different pinch gestures that both resemble natural interactions.  

\vspace{-0.05in}
\subsection{Touchscreen Typing}
\begin{figure}[h]
\begin{center}
\includegraphics[width=85mm]{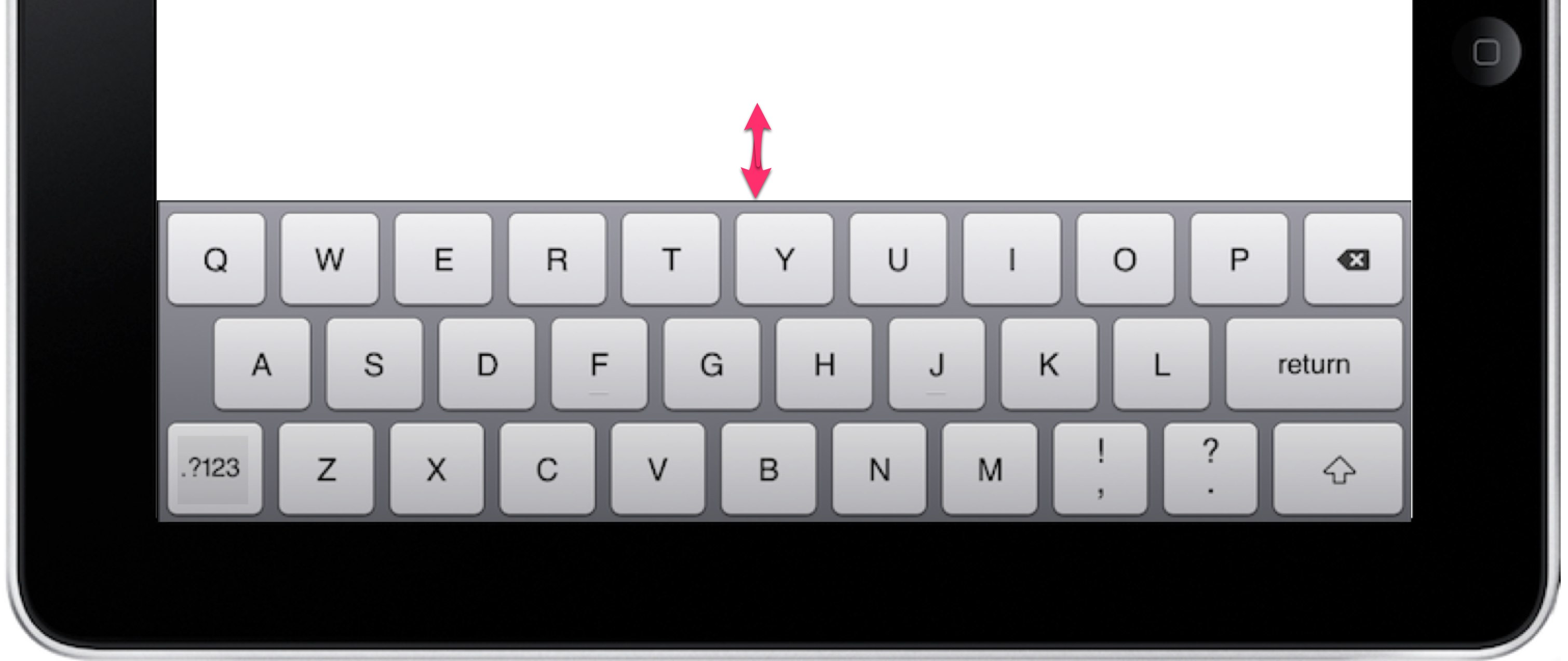}
\caption{By removing the space bar and using a magnetic fingerprint to provide spaces, virtual keyboards could be made 25\% smaller.} 
\end{center}
\end{figure}
Virtual keyboards on tablet devices (in landscape mode) nearly take up half of the screen; which leaves little screen real estate available for applications that rely on typing, such as word processors. 
The spacebar is the most frequently used key but nearly takes up an entire row. We developed a simple text editor application that uses a smaller custom keyboard with no spacebar and only three rows of keys using a minor rearrangement of keys. 
With a magnetic fingerprint attached to the thumb users can activate the space bar by briefly tapping the screen or even using a wireless flick gesture. Backspace could be activated using a magnetic fingerprint on the other thumb.


\vspace{-0.1in}
\section{Discussion and Future Work}

{\it Scalability.} Our experiments reveal that our technique is limited to using up to two magnetic fingerprints at the same time but magnetic fingerprints cannot be chorded. This limitation is due to magnetic interference, as current magnetometers are unable to resolve the complex magnetic fields that appear between using multiple magnets. Despite this limitation magnetic fingerprint significantly increases touch vocabulary. 
Without chording up to 13 unique (multi) touch events can be defined for two magnetic fingerprints and up to 9 unique (multi) touch events for using one magnetic fingerprint, which nearly doubling (one) or triplesl (two) the current touch vocabulary of mobile touchdevices.  

 {\it Comparison.} 
We compare our results to a number of closely related approaches. A precise comparison is difficult due to differences in the increase in touch vocabulary and experimental conditions. Our approach offers a slightly smaller increase in input space as Microrolls \cite{Roudaut:09} (16 micro gestures, 95\% accuracy) but offers a slightly higher accuracy. Thumbrock \cite{Bonnet:13} offers a similar accuracy (96\%) but only adds a single thumb micro gesture. In general, microgestures are a slower method of input  than our approach \cite{Hinckley:05}. Unlike micro gestures, our technique does require a small amount of user instrumentation. Tapsense  \cite{Harrison:2011a}  achieves an accuracy of 95\% and can distinguish four different types of finger touches. Gripsense \cite{Goel:12} identifies three levels of finger pressure with 95\% accuracy. We achieve an accuracy of 96\% for three different types of fingers, but our technique can be used to augment multitouch input (no chording). Tapsense also requires an external acoustic sensor and per user calibration. 
Other related approaches \cite{Boring:12,Benko:06,Hinckley:11} do not report accuracy. 
Looking at other factors, Magnetic fingerprints do not require any external sensors or any form of user calibration and it is low-cost with neodymium magnets retailing for less than \$0.10. 

{\it Limitations.} Finger instrumentation is used in a number of novel input techniques \cite{Weiss:11,Yang:12, Chan:13}. Our approach uses existing sensors and requires the smallest amount of user instrumentation (an untethered small magnet) which seems like a reasonable usability tradeoff to achieve a significant increase in input space with a high accuracy. 
A current limitation of magnetic fingerprints is that false recognitions may occur when a user does not fully retract their magnetized finger if that finger is not involved in the gesture provision. We did not observe this interference in our first study for using a single magnetic fingerprint, but this problem manifested itself only when using two and larger sized magnets on the tablet sized touchscreen, as due to the larger screen users had to reach over the magnetometer to provide input. We observed that some users already retract those fingers not involved in input especially when using smartphone based devices, so this seems a feasible requirement.  
Individual classifiers could be used to mitigate false recognitions, which may depend on the size of the user's hand or the position of their fingers, but this would limit the general applicability of our technique. To keep conditions balanced between trials, our experiments were performed with the device lying flat on the table. In preliminary trials, we did not observe a difference in performance when holding the device in the hand. 

{\it Future research.}  A cube shaped magnet may allows for defining six unique magnetic fingerprints based on the available orientations of the magnet. It may be challenging however to fit this magnet on a finger and it may be harder to classify magnetic fingerprints. To optimize form factor we will explore the use of  magnetic nail polish though this may not generated a strong enough magnetic field. 
Our study demonstrated that using larger magnets allows our technique to be used on tablets, though false recognitions may increase.
To circumvent this problem, we will explore bimanual use with the hand closest to the magnetometer equipped with a smaller magnet and the other hand with larger magnet. 

 
\section{Conclusion}
This paper presents magnetic fingerprints a mobile input technique that instruments the user's finger with a small non-obtrusive magnet. By incorporating sensing information from the device's magnetometer, touches made with a magnetic fingerprint can be distinguished from a non-instrumented finger. 
Unlike existing magnetic input technique, magnetic fingerprints take advantage of the rich vocabulary offered by multitouch input. Magnetic fingerprints are low-cost, require no user calibration and can implemented on current touchscreen devices. 
User studies investigate the accuracy, scalability and limitations of this technique, which found that magnetic fingerprints cannot be chorded.  
Useful applications of magnetic fingerprints are presented.

%
%
%
%
%

\footnotesize

\balance

\bibliographystyle{acm-sigchi}

\end{document}